\documentclass[aps,prl,superscriptaddress,twocolumn,showpacs,noeprint]{revtex4-2}

\usepackage{qcircuit}

\usepackage[normalem]{ulem}
\usepackage{graphicx,amsfonts,times,bm,amsmath,amssymb,verbatim,ifthen,braket,color,array,natbib,bm,soul}
\usepackage{array}
\usepackage{delarray}
\usepackage{xcolor}
\usepackage{subfigure}
\usepackage{xcolor}
\newcommand{\be}{\begin{equation}}
\newcommand{\ee}{\end{equation}}
\newcommand{\bea}{\begin{eqnarray}}
\newcommand{\eea}{\end{eqnarray}}

\newcommand{\renyi}{Rényi }
\newcommand{\Tr}{\mathrm{Tr}}

\usepackage{hyperref}
\usepackage{zx-calculus}

\begin{document}

\title{
Nonunitary gates using measurements only
}

\author{Daniel Azses}
\affiliation{School of Physics and Astronomy, Tel Aviv University, Tel Aviv 6997801, Israel}

\author{Jonathan Ruhman}
\affiliation{Department of Physics, Bar-Ilan University, 52900, Ramat Gan, Israel}
\affiliation{Center for Quantum Entanglement Science and Technology, Bar-Ilan University, 52900, Ramat Gan Israel}

\author{Eran Sela}
\affiliation{School of Physics and Astronomy, Tel Aviv University, Tel Aviv 6997801, Israel}

\begin{abstract}
Measurement-based quantum computation (MBQC) is a universal platform to realize unitary gates, only using measurements which act on a pre-prepared entangled resource state. By deforming the measurement bases, as well as the geometry of the resource state, we show  that  MBQC circuits always transmit  and act on the input state but generally realize nonunitary logical gates. In contrast to the stabilizer formalism which is often used for unitary gates, we find that ZX calculus is an ideal computation method of these nonunitary gates. As opposed to  unitary gates, nonunitary gates can not be applied with certainty, due to the randomness of quantum measurements. We maximize the success probability of realizing nonunitary gates, and discuss applications including imaginary time evolution, which we demonstrate on a noisy intermediate scale quantum device.
\end{abstract}

\maketitle
\noindent\emph{Introduction---} Unitary operations are fundamental for quantum computation from its perception, but a growing interest was recently drawn to the advantages in implementing non-unitary operations directly~\cite{terashima2005nonunitary}, particularly 
via  quantum measurements. This includes quantum steering~\cite{cavalcanti2016quantum,uola2020quantum,roy2020measurement}, measurement induced entanglement transitions~\cite{li2018quantum,PhysRevX.9.031009,chan2019unitary,fux2023entanglementmagic}, as well as applications in quantum computer science for solving NP-complete problems~\cite{abrams1998nonlinear}, block  encoding~\cite{harrow2009quantum,camps2022fable,biswas2023noiseadapted,leadbeater2023nonunitary}, imaginary time evolution~\cite{motta2020determining,sun2021quantum,grundner2023complex} and Lindbladian dynamics simulations~\cite{watad2023variational,lin2021real}.

\begin{figure}[t]
\centering
\includegraphics[width=0.8\linewidth]{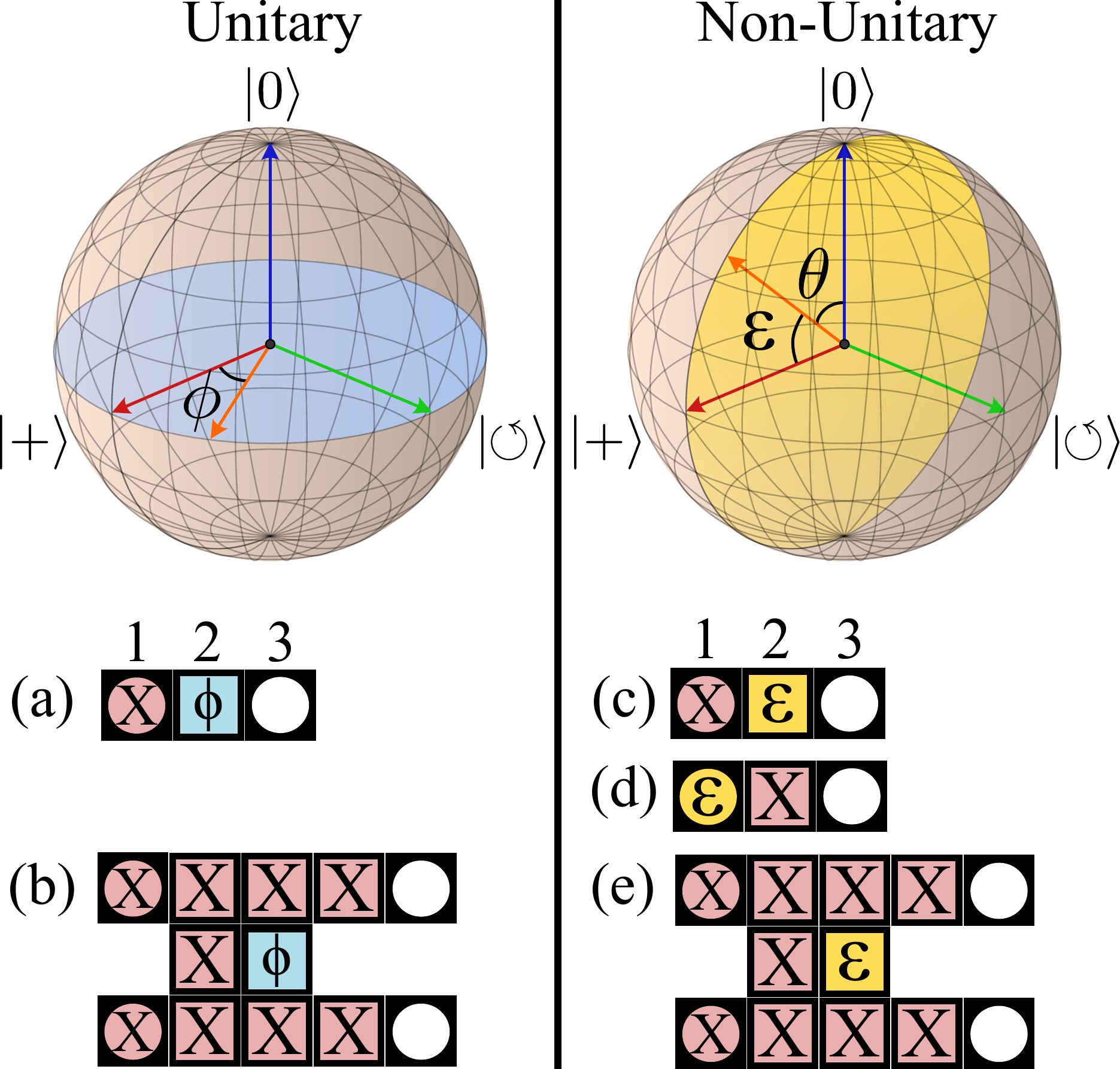}
\caption{\label{fig_mbqc_patterns} Elementary unitary gates using MBQC patterns~\cite{raussendorf2003measurement}. The input and output qubits are denoted with circles. All except output qubits are measured in specified bases. (a,b): Unitary gates implemented by measurements in the $xy$ plane. (c,d,e): Nonunitary gates created by tilting the measurement axis away from the $xy$ plane.}
\end{figure}

In this work we apply measurement-based quantum computation~\cite{raussendorf2001one,raussendorf2003measurement,nautrup2023measurementbased} (MBQC)  to realize nonunitary gates. MBQC proceeds by entangling an input state (left qubits in Fig.~\ref{fig_mbqc_patterns}) with an entangled resource state, typically a cluster state, followed solely by measurements in specified bases, thereby manipulating and propagating quantum information into the output [right qubits in Fig.~\ref{fig_mbqc_patterns}]. Specific unitary gates are realized using specific measurement patterns as exemplified  in Fig.~\ref{fig_mbqc_patterns}(a,b). What happens to these gates as one deforms the measurement bases? More generally, what is the effective action of a MBQC circuit for an arbitrarily selected geometry, e.g. with a different number of input and output gates?

We elucidate that general MBQC patterns act logically on the input quantum information either as unitary or nonunitary gates. While the stabilizer formalism is most adequate for unitary MBQC gates~\cite{raussendorf2003measurement}, we identify ZX calculus~\cite{coecke2008interacting} along with its local ``rules", as the natural approach to deal with non-unitary gates, 
whose elementary building blocks (termed ``spiders") are generally nonunitary. We identify elementary nonunitary gates, which together with unitary gates, form a universal set~\cite{terashima2005nonunitary}. As compared to the quantum circuit model, in which nonunitary gates require ancilla qubits, in MBQC, the latter form part of the resource state. 

Despite of the random nature of quantum measurements, on which MBQC relies, unitary gates can be applied with certainty~\cite{raussendorf2003measurement}. In contrast, there is a fundamental limitation to apply  nonunitary gates deterministically~\cite{terashima2005nonunitary}, which is linked with the randomness and irreversibility of quantum measurements. In the context of measurement induced phase transitions, for example, this is linked with the postselection problem~\cite{nahum2017quantum,li2018quantum}. We find that one can enhance the success probability, compared to that of a single measurement, without losing the logical information. We show how one can saturate the success probability to the maximally allowed one~\cite{terashima2005nonunitary} using a feedback protocol. 

As an  application, we realize an imaginary time evolution protocol, and test it both using matrix product states and on an noisy intermediate-scale quantum (NISQ) device. We also characterize non-unitary gates using an ``operator entanglement"~\cite{zanardi2001entanglement,prosen2007operator,jonay2018coarsegrained,nahum2021measurement}, which measures the entanglement ``in time" between the input and output states, and is independent of the input state. As opposed to recent studies of the operator entanglement of reduced density matrices~\cite{Rath_2023}, our operator entanglement does not assume any spatial bipartition. We demonstrate on a NISQ device a measurement of the operator entanglement using well-known methods.

\noindent
\emph{From unitary to nonunitary MBQC gates---} The difference between unitary and nonunitary MBQC gates can be illustrated via a 1-qubit gate as in Fig.~\ref{fig_mbqc_patterns}(a,c).
Consider a three-qubit cluster state $|\psi_{c}\rangle$, which is the unique common $+1$ eigenstate of the stabilizers $\{ X_1 Z_2, Z_1 X_2Z_3, Z_2X_3 \}$. Now consider the two qubit state $| \psi_{1,3} \rangle$ obtained by measuring the middle qubit (2) and projecting it into the $+1$ eigenstate of $\hat{n} \cdot \vec \sigma$, with, $| \psi_{1,3} \rangle_{\hat{n} } = \hat{\Pi}^{(2)}_{\hat{n} } |\psi_{c} \rangle / \sqrt{\langle \psi_{c} |\hat{\Pi}^{(2)}_{\hat{n} }| \psi_{c} \rangle }$ and $ \hat{\Pi}^{(i)}_{\hat{n} }  = (1+ \vec{\sigma}_i \cdot \hat{n})/2$, and then discarding qubit 2. Consider three cases: (i) If $\hat{n}=\hat{x}$ then the resulting state is the $+1$ eigenstate of (or ``is stabilized by") $Z_1 Z_3$ and $X_1 X_3$, which is the Bell state $|\psi_{1,3} \rangle_{\hat{x} }   = (|00\rangle+|11\rangle)/\sqrt{2}$. It has an  entanglement entropy $S_{{\rm{EE}}}=\log 2$ between qubits 1 and 3. (ii) For any measurement in the $xy$ plane, $\hat{n}_\phi=\hat{x} \cos \phi  + \hat{y} \sin \phi$ one obtains the state $|\psi_{1,3} \rangle_{\hat{n}_\phi } $  stabilized by $Z_1 U_3 Z_3 U_3^\dagger$ and $X_1 U_3 X_3 U_3^\dagger$ where $U_3= e^{i (\phi/2) X_3}$. This state is equivalent to the above Bell state up to a unitary transformation $U=U_3$ acting on qubit 3, and hence it also has an entanglement of $\log 2$. The presence of a Bell pair between input and output in  cases (i,ii) gives the ability to perform unitary MBQC. The construction of Ref.~\cite{raussendorf2003measurement} and theorem 1 therein state that, by inserting an arbitrary input state $| \psi_1 \rangle$ into qubit 1, entangling it with the rest of the cluster state, measuring it in the $\hat{x}$ direction, and finally measuring qubit $2$, these pair of stabilizer equations imply that the state qubit (3) is $|\psi_3 \rangle = U | \psi_1 \rangle$. Thus, quantum information is wired through the chain, along with the action of $U$. However, for (iii) $\hat{n} = \hat{z}$ one obtains the common $+1$ eigenstate of $X_1$ and $X_3$, $|\psi_{1,3} \rangle_{\hat{z} } =|++\rangle$, which is unentangled. Generally, for $\hat{n}_{\theta,\phi} =\hat{z} \cos \theta +\sin \theta (\hat{x} \cos \phi + \hat{y} \sin \phi)$ one obtains a state whose entanglement $S_{{\rm{EE}}}=-p\log p -(1-p) \log (1-p)$ with $p=\sin^2 \frac{\theta}{2}$ varies between $\log 2$ and $0$ as $\theta$ varies from $\pi/2$ to $0$. Although the assumptions of theorem 1 of Ref.~\cite{raussendorf2003measurement} do not apply, we will use this measured cluster state for MBQC.

Generalizing to any MBQC pattern with $n_I$ input and $n_O$ output qubits, by measuring all the remaining qubits of the cluster state, the resulting state is Schmidt-decomposed as
\be
\label{eq:state}
|\psi_{I,O} \rangle = \sum_{\alpha=1}^{2^{{\rm{min}}(n_I,n_O)}} \mu_{\alpha} | \alpha;I \rangle \otimes  | \alpha; O \rangle,
\ee
having entanglement entropy  $S_{{\rm{EE}}}=-\sum_\alpha |\mu_\alpha|^2 \log |\mu_\alpha|^2$. Now consider the {\emph{operator}} 
\be
\label{eq:N}
\hat{N} = \sum_{\alpha=1}^{2^{{\rm{min}}(n_I,n_O)}} \mu_{\alpha}   | \alpha; O \rangle  \langle \alpha;I  | ,
\ee
with  normalization $\sum_\alpha |\mu_\alpha|^2=1$; Like $| \psi_{I,O}\rangle$, $\hat{N}$ is defined up to a complex phase.  This operator is our central quantity of interest. $S_{{\rm{EE}}}$ describes its operator entanglement entropy~\cite{jonay2018coarsegrained,nahum2021measurement}  $S_{\mathrm{op}}=S_{\mathrm{EE}}$, which now quantifies entanglement  in ``time"
rather than space. This quantity can be directly related to the capacity of the measured channel corresponding to a nonunitary evolution operator~\cite{gullans2020dynamical}. For the above single qubit example, $S_{\mathrm{op}}$ equals $\log 2$ in the unitary case $\theta=\pi/2$ where the quantum information is perfectly transmitted through the circuit, and decreases to zero at $\theta=0$. 

\begin{figure*}[t]
\centering
\includegraphics[width=1\linewidth]{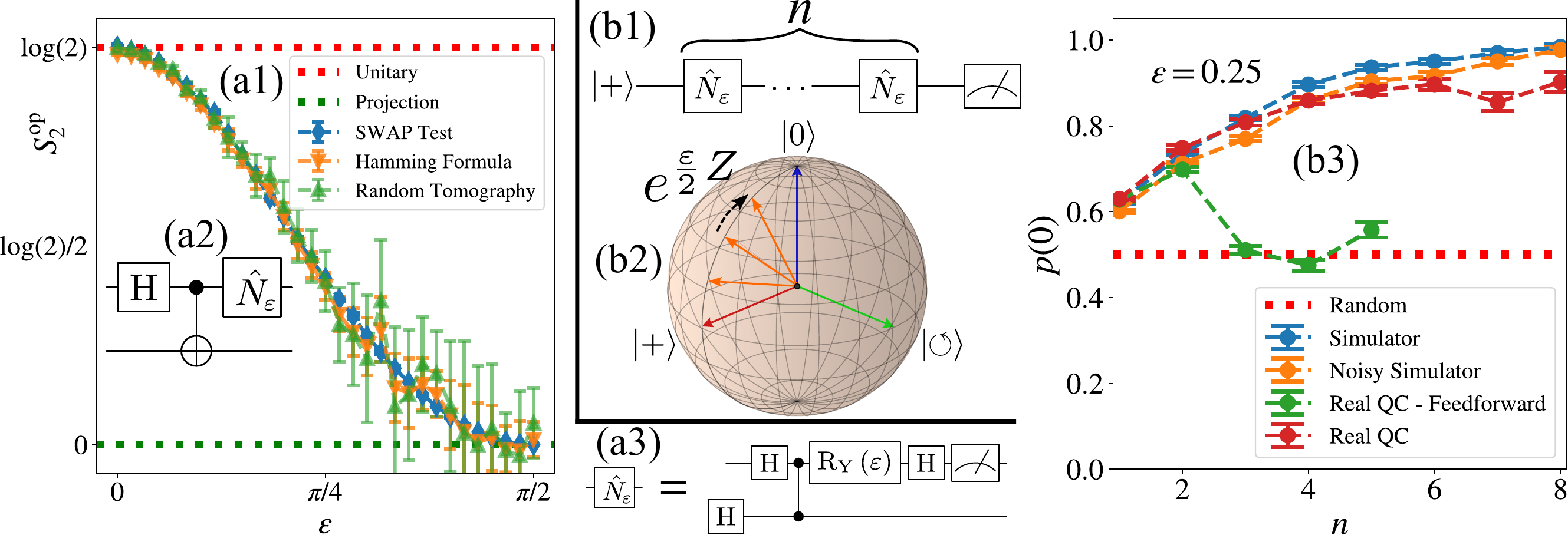}
\caption{\label{fig_imaginary_time} Tests and applications of nonunitary MBQC gates on a NISQ device. (a) We construct $\hat{N}_\varepsilon$ of Fig.~\ref{fig_mbqc_patterns}(d) using a quantum circuit (a3) and measure the second \renyi operator entanglement using either a swap test based on (a2) or using randomized measurements~\cite{SM}. We observe a crossover as function of $\varepsilon=0\to \pi/2$ from a unitary to a projective gate. (b) $n$ applications of $\hat{N}_\varepsilon$ (b1) realizing imaginary time evolution $e^{-\tau H_z}$ with $H_z=-Z$ and $\tau \propto n$ using Trotterization (b2). The probability $p(0)=|\langle 0 | {\rm{out}} \rangle|^2$ displays in (b3) a gradual projection to the ground state of $H_z$, for various implementations, including quantum computers (QC) \cite{SM}. In both (a) and (b) we apply postselection. }
\end{figure*}

Our main practical result is that the state-operator duality in  Eqs.~(\ref{eq:state},\ref{eq:N}) can be realized in MBQC. Namely, suppose the state Eq.~(\ref{eq:state}) is obtained by a measurement of the ``middle" qubits of the cluster state. The procedure is to 
prepare a desired input state $| \psi_I \rangle$, and entangle it with the  cluster state using controlled-Z gates on all nearest neighbors in the desired geometry; By projectively measuring the input gate in the $\hat{x}$ direction, obtaining $|+\rangle$, and  measuring the middle qubits in the specified bases defining $\hat{N}$, yields the output state
\be
|\psi_O \rangle = \hat{N} | \psi_I \rangle /\sqrt{\langle \psi_I | \hat{N}^\dagger \hat{N} | \psi_I \rangle}.
\ee
While this statement seems natural, it requires  proof.
Deforming MBQC away from the realm of unitary gates requires modified methods. The usual stabilizer formalism is not immediately suitable  since the measured state is no longer a stabilizer state nor locally unitary equivalent to it. More general methods are based on matrix product states~\cite{else2012symmetry, fechisin2023noninvertible}, and stress that the computational power relies on a symmetry protected topological feature of the resource state \cite{raussendorf2017symmetry,Raussendorf2022}, and on its internal symmetry structure of entanglement \cite{azses2020symmetry,paszko2023edge}. Here, we apply ZX calculus which is specifically suitable to deal with nonunitary maps. It describes non unitary gates with simple diagrams, and its ``rules” are local
and easy to apply. One of the key applications of ZX calculus~\cite{coecke2008interacting} is a framework for MBQC~\cite{mcelvanney2022complete} for unitary gates. ZX calculus has a symmetry of exchanging the direction of time, which provides the essence of the proof of the above state-operator duality~\cite{SM}. 

\emph{Examples---} Consider the 3 qubit measurement pattern in Fig.~\ref{fig_mbqc_patterns}(c) where the input qubit is measured in the $\hat{x}$ direction, thereby projecting it either into the $+1$ or $-1$ eigenstate of $X_1$. We denote this measurement outcome by $s_1=0,1$. We then perform a measurement of $\hat{n}_{\theta} \cdot \vec{\sigma}$ on qubit 2, with outcome $s_2=0,1$. The resulting nonunitary gate depends on the measurement outcomes ${\mathbf{s}}= \{ s_1 , s_2, \dots\}$, and we denote it as $M_{{\mathbf{s}}}$. We select to normalize our MBQC gates like generalized positive operator-valued measurements (POVM) $\sum_{{\mathbf{s}} } M_{{\mathbf{s}}}^\dagger M_{{\mathbf{s}}}=1$, such that their probability of occurrence is $p_{{\mathbf{s}}} = \langle M^\dagger_{ {\mathbf{s}} } M_{ {\mathbf{s}} } \rangle$.  Skipping the algebra~\cite{SM} we obtain
\be
M_{s_1,s_2}^{\rm{Fig.~1c}}=\frac{1}{\sqrt{2}} H M_{s_2} H Z^{s_1},
\ee
where 
\be
M_0= \frac{1}{\sqrt{1+a^2}}\begin{pmatrix}
a & 0 \\
0 & 1
\end{pmatrix},~~~M_1= \frac{1}{\sqrt{1+a^2}}\begin{pmatrix}
1 & 0 \\
0 & -a
\end{pmatrix},
\ee
where $a=\frac{\sin \theta}{1-\cos \theta}$. We define $\varepsilon=\pi/2-\theta$ such that $a=\frac{\cos \varepsilon}{1-\sin \varepsilon}$. The gate $M_0$ is nonunitary except for $\varepsilon=0$ at which $S_{{\rm{op}}}=\log 2$. Small $\varepsilon$ refers to a weakly nonunitary gate with $S_{{\rm{op}}}=\log 2- \frac{\varepsilon^2}{2} + \mathcal{O}(\varepsilon^4)$.  As $\epsilon \to \pi/2$ it becomes a projector with $S_{{\rm{op}}} \to 0$.

A single nonunitary gate together with a  controlled-not and all 1-qubit unitary gates,  form a universal set of nonunitary gates~\cite{terashima2005nonunitary}. Nevertheless, it can be convenient to construct other elementary nonunitary gates by short circuits. For example, exchanging the measurements bases  of Fig.~\ref{fig_mbqc_patterns}(c), one obtains Fig.~\ref{fig_mbqc_patterns}(d) which yields the gate
\be
\label{eq:main_gate}
M_{s_1,s_2}^{\rm{Fig.~1d}}=\frac{1}{\sqrt{2}} X^{s_2} M_{s_1} .
\ee
It provides for $\varepsilon \to \pi/2$ a  projector in the $Z$ basis  [rather than $X$ basis as in  Fig.~\ref{fig_mbqc_patterns}(c)]. We also give an example of a nonunitary 2-qubit gate in Fig.~\ref{fig_mbqc_patterns}(e), which  is given by~\cite{SM} 
\be
M_{\mathbf{s}}^{\rm{Fig.~1e}} \propto  \left( I -\tan((\varepsilon+s\pi)/2) X_1 X_2 \right) \mathrm{SWAP},
\ee
where $s$ is the measurement result of the qubit marked with $\varepsilon$, while all other qubits are postselected to $|+\rangle$. This is a 2-qubit non-unitary gate representing a non-ideal simultaneous measurement of $X_1X_2$, which becomes completely inefficient at $\varepsilon = 0$, or projective at $\varepsilon=\frac{\pi}{2}$.

\noindent \emph{Testing MBQC nonunitary gates on NISQ devices}---We exemplify in Fig.~\ref{fig_imaginary_time} the application of our nonunitary MBQC gates on a NISQ device by focusing on the gate $\hat{N}_\varepsilon = \sqrt{2} M_{0,0}^{\rm{Fig.~1d}}$. In Fig.~\ref{fig_imaginary_time}(a) we measure  the operator entanglement. For this purpose, we create two copies of a two-qubit state obtained by preparing a Bell state $(|00\rangle+|11\rangle) / \sqrt{2}$ and acting with the gate $\hat{N}_\varepsilon$ on one of the qubits. $S_{\rm{op}}$ is the entanglement between the two qubits~\cite{SM}. The second \renyi operator entanglement $S_{\rm{op}}^{(2)}=-\sum_\alpha \log \mu_\alpha^4$ can be measured~\cite{johri2017entanglement} by a swap test of the two copies or randomized measurements~\cite{SM}. We can see that the operator entanglement crosses from $\log 2$ in the unitary case $\varepsilon=0$ to $0$ in the projective measurement limit $\varepsilon \to \pi/2$. In the latter limit the circuit has zero capacity to transmit quantum information.

Next we discuss imaginary time evolution. One can simulate real time evolution of many-body problems, e.g. the Fermi Hubbard model, by constructing the unitary operator $U=e^{-i t H}$ using unitary gates, where time $t$ translates in MBQC into measurement angle in the $xy$ plane~\cite{PhysRevResearch.4.L032013}. For example, the two qubit gate in Fig.~\ref{fig_mbqc_patterns}(b) allows to implement the gate~\cite{SM} $e^{-i(\phi/2)X_1 X_2}$, which describes real time evolution $t \propto \phi$ under an Ising Hamiltonian $X_1 X_2$. Combining with single qubit gates, one can simulate general Hamiltonians. Using rotated bases of measurements, our nonunitary two qubit and single qubit gates can simulate imaginary time evolution $e^{-\tau H}$ where $\tau \propto \varepsilon$. As opposed to Ref.~\cite{mao2023measurement} which averages shot-to-shot measurements to obtain the imaginary time evolution using conditional unitaries, our simulation obtains the evolution for any finite time $\tau$ by employing post-selection and measurements only.

In Fig.~\ref{fig_imaginary_time}(b) we realize a 1-qubit imaginary time evolution for imaginary time $\propto \varepsilon$ using the nonunitary gate $\hat{N}_\varepsilon$. By concatenating $n$ such gates, we obtain a gate $ e^{-\tau H_z}$ with imaginary time $\tau= \frac{n}{2}\log a$ and Hamiltonian $H_z=-Z$. Taking for example an input  state in the equator of the Bloch sphere, $ |+ \rangle $, imaginary time evolution gradually tilts it towards the $|0\rangle$ state at the north pole \cite{SM}. In the above simulations, with multiple applications of nonunitary gates, we performed a postselection of the measurement outcomes. This becomes a serious problem in a circuit with many nonunitary gates.

\noindent\emph{Minimizing the post-selection problem}---The measurement results $s_1,s_2,\dots$ are random. Each outcome corresponds to a different wave function $|\psi_{I,O} \rangle$ and thus to a different MBQC gate. In unitary MBQC, it turns out that the only consequence of randomness, is that on top of the desired gate the random outcomes result in an additional unitary Pauli gate, called byproduct operator, which can be easily inferred using classical tomography. 
In contrast, nonunitary gates  cannot be applied deterministically due to a fundamental  limitation~\cite{abrams1998nonlinear}. 
To appreciate the challenge we first emphasize the pitfall of a naive approach. Namely, one could try to correct an undesired measurement  in the $\hat{n}$-th basis (with eigenstates $|\pm_{\hat{n}} \rangle$) by applying a spin flip gate $| -_{\hat{n}} \rangle \langle  +_{\hat{n}}| 
 + |  +_{\hat{n}} \rangle  \langle -_{\hat{n}}|$ on the measured qubit. However this is not equivalent to directly obtaining the desired measurement outcome. To see this, consider the 3-qubit examples (i,ii,iii) above. The initial cluster state can be written in the $\hat{n}$-th basis for qubit 2 as $|\psi_c \rangle= a_+ |\psi_{1,3} \rangle_{\hat{n}}  \otimes |+_{\hat{n}} \rangle +a_-|\psi_{1,3}' \rangle_{\hat{n}} \otimes |-_{\hat{n}} \rangle$ where $a_\pm$ are coefficients. The desired state after measurement is $|\psi_{1,3} \rangle_{\hat{n}} \otimes |+_{\hat{n}} \rangle $. However, the undesired measurement outcome followed by a flip of qubit 2 gives  $|\psi_{1,3}' \rangle_{\hat{n}} \otimes |+_{\hat{n}} \rangle $. 
 
In what follows, we outline a protocol designed to enhance the likelihood of implementing a desired nonunitary operation. 
However, employing such a protocol will never deterministically lead to the intended operation. Indeed, Ref.~\onlinecite{abrams1998nonlinear}, shows that such an  ability implies the power to solve {\bf{NP}}-complete problems in polynomial time \cite{terashima2005nonunitary}, in contradiction to the substantial hypothesis that {\bf{NP}} $\nsubseteq$ {\bf{BQP}} \cite{aaronson2009bqp}. Moreover, such an ability enables to circumvent the post-selection problem that hinders the experimental observation of the measurement-induced phase transition~\cite{PhysRevX.9.031009} by repeatedly simulating a specific quantum trajectory. The absence of a deterministic protocol raises the question regarding the bound on the probability to apply a certain non-unitary operation and how to saturate this bound. 

When we attempt to realize the gate $\hat{N}_{\varepsilon}$ by the pair of measurements $s_1,s_2$ in Fig.~\ref{fig_mbqc_patterns}(d), we get 4 possible measurement outcomes $(s_1,s_2) \in \{ (0,0),(0,1),(1,0),(1,1)\}$. We declare on $(0,0)$ as successful. As Eq.~(\ref{eq:main_gate}) shows, the outcome  $(0,1)$ is equally successful since it is related to $(0,0)$ by a byproduct operator. However, the cases with $s_1=1$ are unsuccessful. For a generic state $|\psi \rangle=\cos \frac{\beta}{2} |0\rangle + e^{i\varphi}  \sin \frac{\beta}{2} |1\rangle \equiv | \psi_{\beta,\varphi} \rangle$, the success probability is $\sum_{s_2}\langle \psi |M^\dagger_{0,s_2}M_{0,s_2} | \psi \rangle=\frac{a^2 \cos^2 \frac{\beta}{2}+  \sin^2 \frac{\beta}{2} }{1+a^2}$. Can we enhance this probability?

Ref.~\onlinecite{terashima2005nonunitary} determines the maximal success probability of a nonunitary gate $\hat{N}$, see Eq.~(\ref{eq:pmax}). For their argument, it is convenient to use a third normalization convention with unit maximal eigenvalue, $\tilde{N}=\frac{\hat{N}}{{\rm{max}}~ {\rm{eig}} (\hat{N}) }$. Now, they introduce a new pair of POVM operators~\cite{terashima2005nonunitary} 
\be
M^{(c)}_{0}=c  \tilde{N},~~~M^{(c)}_{1}=\sqrt{1-(M^{(c)}_{0})^\dagger M^{(c)}_{0}},
\ee
parameterized by a complex number $c$ satisfying $|c| \le 1$. The idea is that if we can  realize this set of POVMs, then, we will be able to apply the gate $\hat{N}$ with probability $p_1=\langle (M^{(c)}_{0})^\dagger M^{(c)}_{0} \rangle = |c|^2 \langle \tilde{N}^\dagger \tilde{N} \rangle \le 1$. The maximal success probability then is attained at $c=1$, $p_{{\rm{max}}}= \langle \tilde{N}^\dagger \tilde{N} \rangle=\frac{\langle \hat{N}^\dagger \hat{N} \rangle}{ {\rm{max}} ~{\rm{eig}} ~(\hat{N}^\dagger \hat{N}) }$. For $\hat{N}=M_0(a)$ and $|\psi_{\beta,\varphi} \rangle$ this gives 
\be
\label{eq:pmax}
p_{{\rm{max}}}=\cos^2(\beta/2)+ a^{-2} \sin^2 (\beta/2).
\ee
Our 3-qubit MBQC architecture realizes these generalized measurements only with $c= \frac{a}{\sqrt{1+a^2}} \le 1$. 

\begin{figure}[t]
\centering
\includegraphics[width=\linewidth]{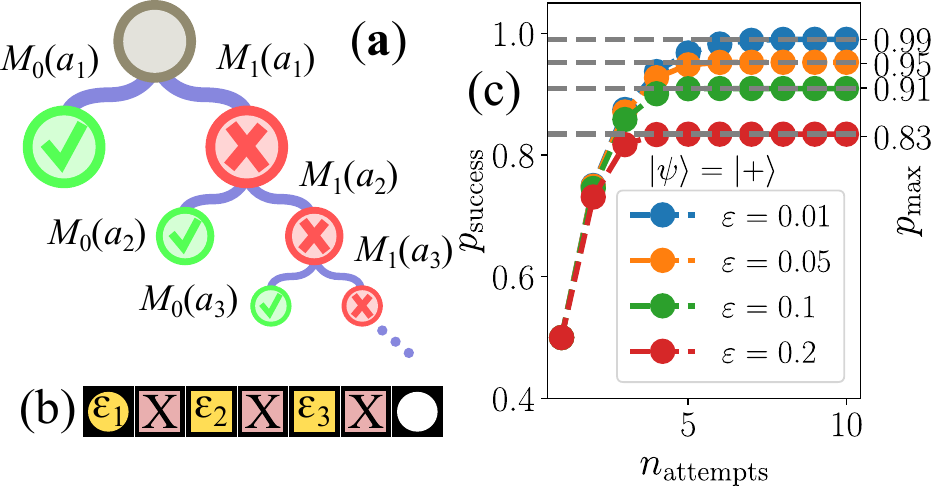}
\caption{\label{fig:mes_prob_success}(a) Measurement scheme as a decision tree of the MBQC pattern in (b) for $3$ branches. Each branch indicates the possible measurement results. (c) Probability of success of correcting undesired measurement outcomes after $n$ attempts according to Eq.~\ref{se:p_success_n} }
\end{figure}

As displayed in  Fig.~\ref{fig:mes_prob_success}(b), using a chain of qubits, we now describe a feedback protocol which enhances the success probability all the way to this maximal allowed value. Here we focus only on the randomness associated with the measurements with angles $\varepsilon_i$ in the $xz$ plane and postselect the $X$ measurements - the latter post-selection is easy to avoid by using conditional gates as in standard MBQC. Our task is to apply the gate $M_0(a)$ for some desired  $a$ denoted $a=a_1$ which corresponds to $\varepsilon=\varepsilon_1$. 
The success probability after one attempt is $p_{1}=\langle M_0^\dagger(a_1) M_0 (a_1) \rangle $. However the undesired operator $M_1(a_1)$ is  applied with probability $1-p_1$. In the latter case we can try to correct it by applying a second gate selected to be $M_0(a_{2})$ with $a_{2}=-a^2$, satisfying $M_0(a)=M_0(a_2) M_1(a)$, which will undo the random mistake of the former measurement. This is performed by the same MBQC circuit in Fig.~\ref{fig_mbqc_patterns}(d), with different angle $\varepsilon_2$ such that $a_2=a(\varepsilon_2)$, see Fig.~\ref{fig:mes_prob_success}(b). The success probability of the second attempt is $p_{2}=\langle  M_1^\dagger(a_1)  M_0^\dagger(a_{2}) M_0 (a_{2})  M_1(a_1)  \rangle $. Proceeding, the success probability in the $n-$th attempt is then $p_n= \langle  ( \prod_{i=1}^{n-1} M_1^\dagger(a_{i})  )   M_0^\dagger(a_{n}) M_0 (a_{n}) ( \prod_{i=1}^{n-1} M_1(a_{i}) ) \rangle$ with $a_n=a^{2^{n-1}} (-1)^{n-1}$. Using $\prod_{i=1}^{n-1} a_{i} =(-1)^{n} a^{2^{n-1} -1}$, the success probability in the $n-$th attempt quickly decreases with $n$ as $p_n=p_{{\rm{max}}} \frac{a^2-1}{a^{2^n} - a^{-2^n}}$, as illustrated in Fig.~\ref{fig:mes_prob_success}(a), and the total success probability after $n$ attempts is 
\be
\label{se:p_success_n}
p_{\rm success} (n)= \sum_{i=1}^{n} p_i= p_{{\rm{max}}} \left(\sum_{i=1}^{n}  \frac{a^2-1}{a^{2^i} - a^{-2^i}}  \right).
\ee
The sum in the second factor tends to unity for $n \to \infty$,  saturating the maximal success probability of Ref.~\cite{terashima2005nonunitary}. 
This is plotted in Fig.~\ref{fig:mes_prob_success}(c) for $\beta=\pi/2$. 
As we can see in Fig.~\ref{fig:mes_prob_success}(c), $p_{\rm success}(n)$ tends to unity for small $\varepsilon$, as $p_{\rm success}(n\to \infty) = 1-2 \sin^2(\beta/2) \varepsilon$. 
In general, $p_{\rm success} (n)$ saturates at a sub-unitary success probability, i.e. there is a finite probability of failure.

\noindent\emph{Summary and Outlook}-- The inclusion of non-planar measurements in the MBQC formalism corresponds to the application of nonunitary operations in the logical level. The most natural interpretation of such operations is obtained within ZX calculus. We demonstrated the application of these nonunitary MBQC gates on a real NISQ machine and we also discussed the application of multi-site non-unitary gates. A specific nonunitary operation cannot be deterministically applied, unlike unitary gates. However, using a Markovian monitor-and-correct protocol the probability  of applying a desired non-unitary operation can be increased, depending on how close the operation is to a perfect projection. Our results open a new path to implement non-unitary operations in quantum information processing, whether directly as a computational tool or to simulate a lossy environment. 

These results naturally lead to the question of whether our monitor-and-correct protocol can reduce the post-selection overhead required to observe the measurement phase transition experimentally, which is an outstanding challenge~\cite{PhysRevX.9.031009,noel2022measurement}. On one hand, our method is useless in the limit of perfect projections, but on the other hand, the entangling gates can be chosen in a way to lower the critical threshold. Furthermore, the interpretation of non-unitary circuits undergoing the measurement phase transition in terms of ZX-calculus diagrams may lead to new insights.

Finally, the computational power of unitary MBQC was found to go beyond the specific cluster state and to rely only on symmetry~\cite{else2012symmetry,Miller_2018,raussendorf2019computationally,qin2023redundant}. It would be interesting to look at the nonunitary gates, and particularly on their entanglement~\cite{Goldstein_2018}, from this symmetry perspective. 

\noindent\emph{Acknowledgments--} D.A. and E.S. acknowledge support from the European Research Council (ERC) Synergy funding for Project No. 951541 and ARO (W911NF-20-1-0013). J.R. was supported by the Israeli Science Foundation Grant No. 3467/21. We thank Mario Motta for fruitful discussions. We acknowledge the use of IBM Quantum services for this work. The views expressed are those of the authors, and do not reflect the official policy or position of IBM or the IBM Quantum team.

\begin{appendix}
\newpage
In this Supplemental Material we provide further
details on (i) ZX calculus for nonunitary gates, 
(ii) Quantum computer implementations, (iii) operator entanglement and (iv) the feedback correction protocols.

\section{ZX calculus for nonunitary gates}
We start by briefly introducing ZX calculus.

{\emph{Definition:}} Green and red spiders are the following linear maps: 
\bea
\begin{ZX}
\leftManyDots{n} \zxZ{\alpha} \rightManyDots{m}
\end{ZX}
&=& |\underbrace{000}_{m}\rangle \langle \underbrace{000}_{n}|+e^{i \alpha} |111\rangle \langle 111|, \\
\begin{ZX}
\leftManyDots{n} \zxX{\beta} \rightManyDots{m} 
\end{ZX}
&=& |\underbrace{+++}_{m}\rangle \langle \underbrace{+++}_{n}|+e^{i \beta} |---\rangle \langle ---|. \nonumber 
\eea
Here information flows from left to right, but ZX calculus is  invariant to this specification. For example, for $n=m=1$ these are single qubit phase gates, e.g. \begin{ZX}
\zxN{} \ar[r]  & \zxZ{\alpha} \arrow[r] & \zxN{}  \end{ZX}  $= |0\rangle \langle 0 | +e^{i \alpha}|1\rangle \langle 1 |$. In general, these linear maps are {\emph{nonunitary}}. A spider acting on the vacuum and returning one qubit ($n=0, m=1$) can be viewed as a state, for example
\begin{ZX}
   \zxZ{} \arrow[r] & \zxN{} \end{ZX} $= |0\rangle +|1\rangle =|+\rangle$.
A spider acting on one qubit and returning a number is a projective measurement, \begin{ZX}
 \zxN{} \ar[r]  &   \zxZ{}  \end{ZX} $= \langle 0 | +\langle 1 |=\langle + |$.
 Likewise, \begin{ZX}
\zxN{} \ar[r]  &   \zxX{} \end{ZX} $= \langle + | +\langle - |=\langle 0 |$.

Universality of ZX calculus states that any linear map can be represented in terms of ZX diagrams, i.e. as a spider web~\cite{coecke2008interacting}. Interestingly, unitary gates which naively seem elementary, such as CNOT, can be decomposed into nonunitary gates~\cite{coecke2008interacting}. ZX calculus has local {\emph{simplification rules.}} This includes fusion rules of spiders of equal color, \begin{ZX}
\leftManyDots{n} \zxX{\alpha} \zxLoopAboveDots{} \middleManyDots{} \ar[r]
& \zxX{\beta} \zxLoopAboveDots{} \rightManyDots{m}
\end{ZX}
$= $
\begin{ZX}
\leftManyDots{n} \zxX{\alpha+\beta} \rightManyDots{m}
\end{ZX}. Additionally, one defines the Hadamard gate explicitly as a yellow square
\begin{ZX}
\zxN{} \arrow[r] & \zxN{} \arrow[r,H] &   \zxN{} \arrow[r] & \zxN{} 
\end{ZX} $=$ \begin{ZX} \zxN{} \arrow[r] &   \zxZ{\pi/2} \arrow[r] & \zxX{\pi/2} \arrow[r] &\zxZ{\pi/2} \arrow[r] &\zxN{} \end{ZX}, which is also notated by a blue dashed line \begin{ZX}
\zxN{} \arrow[r] & \zxN{} \arrow[r,H] &   \zxN{} \arrow[r] & \zxN{} 
\end{ZX} $=$ \begin{ZX}
\zxN{} \arrow[r,blue,dashed] & \zxN{} \arrow[r,blue,dashed] &   \zxN{} \arrow[r,blue,dashed] & \zxN{} 
\end{ZX}. It combines to a controlled-Z gate when inserted between green spiders $\mathrm{CZ} = 
\zx{
\zxN{} \ar[r] & \zxZ{} \ar[d,H] \rar & \zxN{}  \\
\zxN{} \rar & \zxZ{} \rar & \zxN{} \\
} = \zx{
\zxN{} \ar[r] & \zxZ{} \ar[d,blue,dashed] \rar & \zxN{}  \\
\zxN{} & \zxN{} \ar[d,blue,dashed] & \zxN{}  \\
\zxN{} \rar & \zxZ{} \rar & \zxN{} \\
}$.

One of the key applications of ZX calculus~\cite{coecke2008interacting} is a framework for MBQC which was discussed in detail~\cite{mcelvanney2022complete} for unitary gates. We outline now how unitary gates in MBQC are described by ZX calculus. Consider the measurement pattern in Fig.~\ref{fig_mbqc_patterns}(c) with vanishing angles (measurements in the $x$-axis) on three qubits. Namely, we prepare the cluster state   $|cs\rangle= \zx{\zxZ{} \ar[r] & \zxZ{} \ar[d,H] \rar & \zxN{}  \\
\zxZ{} \rar & \zxZ{} \ar[d,H] \rar & \zxN{} \\
\zxZ{} \rar & \zxZ{} \rar & \zxN{} \\
} $ and then measure the input (top) and middle qubit in the $X$ basis. We denote the measurement results by $s_{1,2}=0,1$. This is the simplest teleportation circuit:
\be
\label{eq:mbqc_nutshell}
\zx{\zxN{} \ar[r] & \zxZ{} \ar[d,H] \rar & \zxZ{s_1 \pi}  \\
\zxZ{} \rar & \zxZ{} \ar[d,H] \rar & \zxZ{s_2 \pi} \\
\zxZ{} \rar & \zxZ{} \rar & \zxN{} \\
} = 
\zx{\zxN{} \ar[r] & \zxZ{s_1 \pi} \ar[d,H] \\
& \zxZ{s_2 \pi} \ar[d,H]  \\
& \zxZ{} \rar & \zxN{} \\
}  = 
\zx{\zxN{} \ar[r] & \zxZ{s_1 \pi} \ar[d] \\
& \zxX{s_2 \pi} \ar[d]  \\
& \zxN{} \rar & \zxN{} \\
}  =X^{s_2} Z^{s_1}.
\ee
Here yellow squares are Hadamard gates $(H)$.
What we obtained is teleportation, i.e. the identity gate, up to a unitary operator $U_\Sigma=X^{s_2} Z^{s_1}$ called byproduct operator, that depends on the random measurement outcomes. 

\noindent\emph{Nonunitary gates with ZX calculus---}
From ZX calculus we may technically understand why it is important that measurements of the cluster state in unitary MBQC are always performed in the $xy$ plane. While measurement in the $xy$ plane at an angle $\alpha$ with respect to the $x$ axis is described by unilegged green spiders, tilting the measurement into the $z$-axis involves also red spiders. In the teleportation gate Eq.~(\ref{eq:mbqc_nutshell}) we used the simplification rules to ``fuse" projective measurements (green spides $s_{1,2} \pi$) with unitary gates, to result in an overall unitary gate. This is not the case when tilting the measurement axis of $| cs\rangle$ into the $z-$axis.

Explicitly, we have
\bea
xy{\rm{~measurement:}}& & \begin{ZX} \zxN{} \arrow[r] &   \zxZ{\phi+\pi s} \end{ZX},\nonumber \\
xz{\rm{~measurement:}}& & \begin{ZX} \zxN{} \arrow[r] &   \zxZ{\pi/2} \arrow[r] &   \zxX{\theta+\pi s} \end{ZX}.
\eea
The first equation represents a measurement of a qubit in the $xy$ plane, at angle $\phi$ w.r.t to the $\hat{x}$ axis, yielding an outcome $s=0,1$. In the second equation $\theta = \pi/2-\varepsilon$ is the angle with respect to the $z$ axis. 

A convenient step to incorporate non-$xy$ measurements in MBQC, is
to express projective measurements of red spiders as a linear superposition of green spiders, for example
\bea
\label{eq:yz_measurement}
\begin{ZX}
\zxN{} \arrow[r] &   \zxX{\theta+\pi s} \end{ZX} = \begin{ZX} \zxN{} \arrow[r] &   \zxZ{0} \end{ZX} + e^{i (\theta + \pi s)} \begin{ZX} \zxN{} \arrow[r] &   \zxZ{\pi}
\end{ZX} .
\eea
Each green spider here can be fused into the unitary circuit. Now the gate in Fig.~\ref{fig_mbqc_patterns}(d) can be expressed as 
\be
\label{eq:imaginary_time_step}
\mathcal{N}_{s_1,s_2}=\zx{\zxN{} \ar[r] & \zxZ{} \ar[d,H] \rar & \zxZ{
\pi/2} \rar & \zxX{\theta+\pi s_1}  \\
\zxZ{} \rar & \zxZ{} \ar[d,H] \rar 
&  \zxZ{s_2 \pi}   \\
\zxZ{} \rar & \zxZ{} \rar & \zxN{} \\
} 
= \sum_{\ell=0,1} e^{i \ell (\theta+\pi s_1)} 
\zx{\zxN{} \ar[r] & \zxZ{\pi/2+\ell \pi} \ar[d,H] \\
& \zxZ{s_2 \pi} \ar[d,H]  \\
& \zxN{} \rar & \zxN{} \\
}.
\ee
This becomes $\mathcal{N}_{s_1,s_2}=X^{s_2}  M_{s_1}$ with $X^{s_2}$ being a byproduct operator and 
\be
\label{Ns1s2}
M_{s_1} \propto 
\begin{pmatrix}
1+e^{i(\theta+\pi s_1)} & 0 \\
0 & i-ie^{i(\theta +\pi s_1)}
\end{pmatrix}.
\ee

We note that the gates $M_0$ and $M_1$ form a pair of  positive operator-valued measure (POVM) operators and can be written as 
\be
M_0= \frac{1}{\sqrt{1+a^2}}\begin{pmatrix}
a & 0 \\
0 & 1
\end{pmatrix},~~~M_1= \frac{1}{\sqrt{1+a^2}}\begin{pmatrix}
1 & 0 \\
0 & -a
\end{pmatrix}
\ee
where $a=\frac{\cos \varepsilon}{1-\sin \varepsilon}$. We used the simple  algebra
\bea
\frac{1+e^{i\theta}}{i-ie^{i\theta } } = \frac{\sin \theta}{1- \cos \theta} =\cot \frac{\theta}{2}= \frac{\cos \varepsilon}{1- \sin \varepsilon} \equiv a(\varepsilon),
\eea

\begin{figure}[t]
\centering
\includegraphics[width=0.8\linewidth]{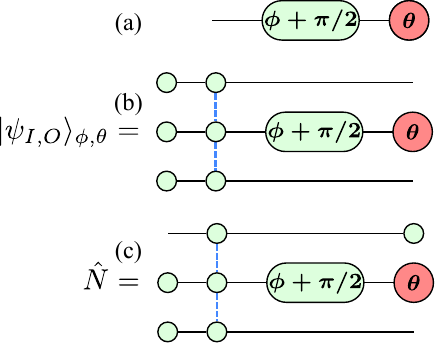}
\caption{\label{fig:app_proof}(a) A general measurement in the polar coordinated $\hat{n} = (\phi,\theta)$ direction on the Bloch sphere. (b) MBQC scheme in ZX calculus that measures the middle qubit in the $\hat{n}$-th direction with outcome $0$. This creates an entangled state between the input $I$ and the output $O$. (c) An equivalent ZX diagram that shows the entangled state as an operator $\hat{N}$. This proves that the entanglement of the state is the same as the operator entanglement of its MBQC action for any measurement outcome.  }
\end{figure}

Their probabilities are $\langle  M_{0,1}^\dagger M_{0,1}  \rangle$. We display an explicit proof of this statement for the 3 qubit state, in Fig.~\ref{fig:app_proof}. Likewise, in Fig.~\ref{fig:app_proof} we illustrate the state-operator duality, connecting Eqs.~(1) and (2) in the main text, using ZX calculus.

 \begin{figure}[t]
\centering
\includegraphics[width=0.8\linewidth]{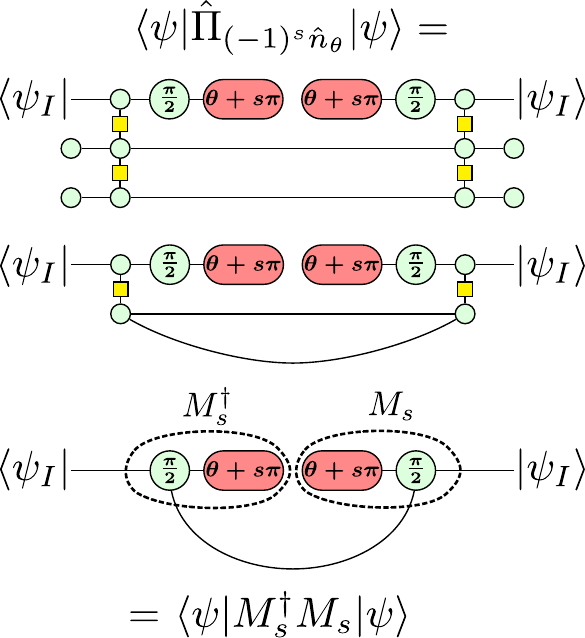}
\caption{\label{fig:prob_proof} Diagramatic proof that the projector of the MBQC state $\ket{\psi}$ encoding the input $\ket{\psi_I}$ has probability $\bra{\psi} M_s^{\dagger} M_s \ket{\psi}$ to measure the first qubit (generally, any qubit) with outcome $s$.}
\end{figure}

\subsubsection{Realizing a two-qubit non-unitary gate}

\begin{figure}
\includegraphics[width=\linewidth]{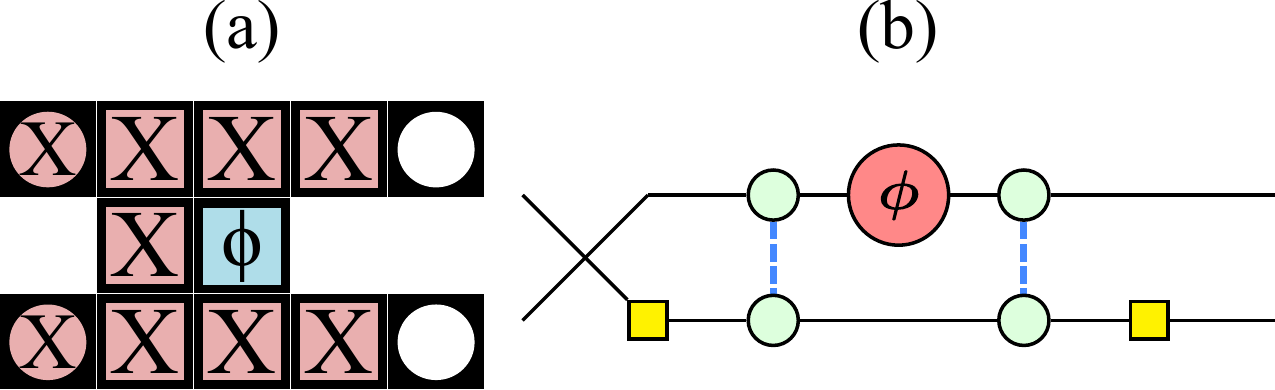}
\caption{\label{fig:2d_grid_scheme} (a) Measurement pattern  used in Ref.~\cite{PhysRevResearch.4.L032013} for a unitary evolution of two qubits. The inputs are the far left and the outputs are the far right qubits. All non-output sites are measured in the $X$ direction except for one site that is measured at an angle $\phi$ in the $xy$ plane. (b) An equivalent simplified ZX diagram obtained by computer calculation. The unitary evolution is $U = e^{-i (\phi/2) X_1 X_2} \mathrm{SWAP}$. When the angle measured is in the $xz$ plane, the operation turns to be a non-unitary one as described in Fig.~\ref{fig:2d_grid_simplification} and the main text.}
\end{figure}

Let us exemplify a nonunitary gate acting on two qubits. First consider a 2D grid with 2 inputs and 2 outputs as in Fig.~\ref{fig:2d_grid_scheme}(a), where $\phi$ is a measurement angle in the $xy$-plane. From now on we assume the  measurement results are $0$. Then the resulting gate is known  to be the unitary~\cite{PhysRevResearch.4.L032013} gate $U = e^{-i \phi/2 X_1 X_2} \mathrm{SWAP}$. This result can also be understood using ZX calculus as in Fig.~\ref{fig:2d_grid_scheme}(b). However, what happens when one measures the special qubit in the $xz$ plane at an angle $\varepsilon$ instead? 

\begin{figure*}
\includegraphics[width=0.97\linewidth]{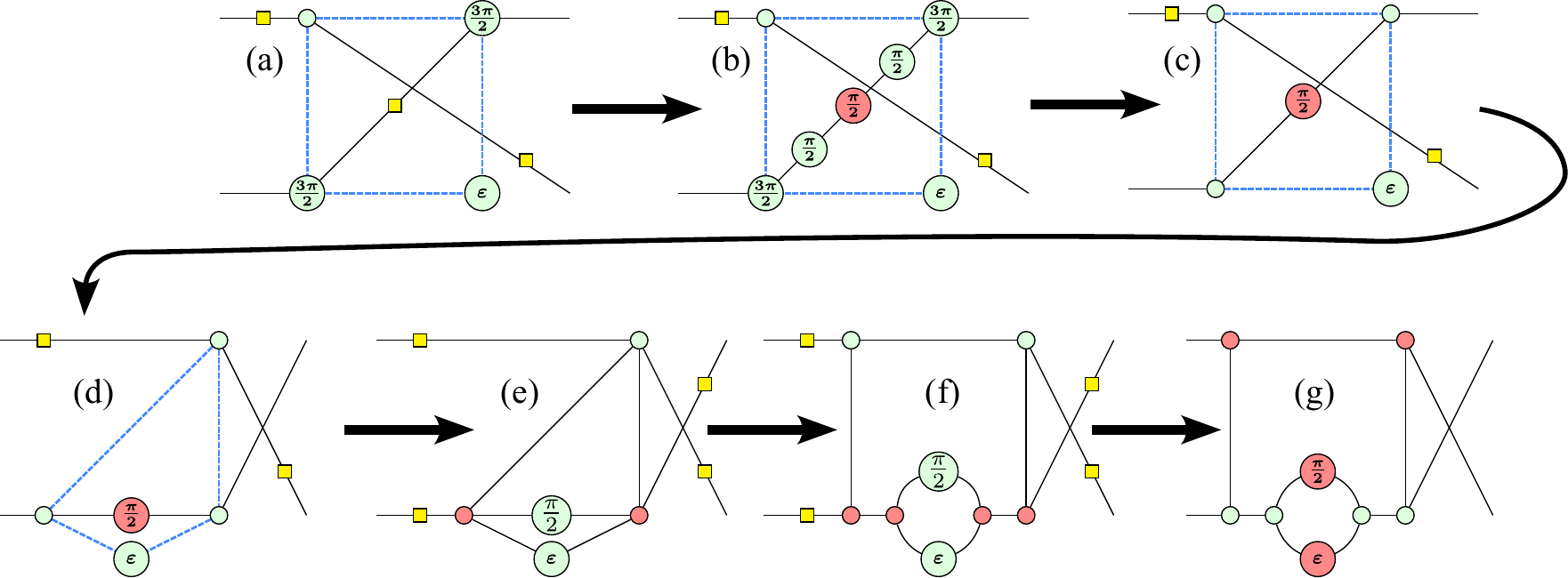}
\caption{\label{fig:2d_grid_simplification}   (a) Computer assisted simplified expression of the MBQC pattern. The computer cannot further optimize the diagram due to its non-unitary nature. (b-g) ZX calculus steps to obtain the simplified   diagram. This shows explicitly the SWAP gate and the control-X gates. The non-unitary blob gate is shown in the 2nd line in sub-figure (g).}
\end{figure*}

For this non-unitary case, we obtain the ZX diagram by a computer simplification using PyZX Python library \cite{kissinger2020Pyzx} and further simplify step-by-step, see Fig.~\ref{fig:2d_grid_simplification}. The only non-unitary part is a bubble on the 2nd qubit:
\begin{ZX}
\zxN{}      & \zxN{}                      & \zxFracX{\pi}{2} \ar[rd,('] & \zxN{}    & \zxN{}  \\
\zxN{} \rar & \zxZ{} \ar[ru,('] \ar[rd,(.]& \zxN{}                  & \zxZ{} \rar & \zxN{}  \\
\zxN{}      & \zxN{}                      & \zxX{\varepsilon} \ar[ru,(.] & \zxN{}    & \zxN{}  \\
\end{ZX}.
This type of gate is manifested by two (unnormalized) equations:
\bea
\ket{0} & \rightarrow &  \ket{0} - \tan(\varepsilon /2) \ket{1} , \nonumber \\
\ket{1} & \rightarrow &  -\tan(\varepsilon /2) \ket{0} + \ket{1} .
\eea
This implements a non-unitary gate on the 2nd qubit 
\be
P = \begin{pmatrix} 
1 & -\tan(\varepsilon/2) \\ 
-\tan(\varepsilon/2) & 1
\end{pmatrix}=I -\tan(\varepsilon/2) X.
\ee

Combining this with the control-$X$ gates from both sides with the non-unitary $P$ in between and the SWAP we get a definite  non-unitary gate. We move the SWAP to act first, and the resulting (unnormalized) action is
\bea
N & \propto & \mathrm{CX}_{1,2} P_1 \mathrm{CX}_{1,2} \mathrm{SWAP} = (I-\tan(\varepsilon/2)X_1 X_2)\mathrm{SWAP} \nonumber \\
&=& \left( \cos\frac{\varepsilon}{2} I-\sin\frac{\varepsilon}{2} X_1 X_2 \right)\mathrm{SWAP}.
\eea

In the case that the measurement results are non-zero one may get additional byproduct operators. For the case that the $\varepsilon$ measurement is non-zero, one transforms $\varepsilon \rightarrow \varepsilon + s\pi$. Hence, the nature of the action is left unchanged.

\section{Quantum computer implementations}

\subsubsection{Quantum computer calibration \& properties}

\begin{table*}
\caption{\label{tab:qubit_params_1} Calibration data for the IBM quantum computer {\it ibm\_brisbane} at November 14, 2023, 02:00 AM, Israel time \cite{IBMQuantum}. This calibration matches the time of running and obtaining the results for the optimized circuit plotted at Fig.~\ref{fig_imaginary_time}(b3). We used 9 qubits and showing the calibration for them only.}
\begin{ruledtabular}
\begin{tabular}{ccccccccc}
Qubit name & Frequency [GHz] & T1 [us] & T2 [us] & Readout error & ID error & $\sqrt{X}$ error & Pauli-X error & Next neighbor ECR error\\
Q0 & 4.7219 & 65.8015 & 33.3981 & 3.17e-02 & 1.0254e-03 & 1.0254e-03 & 1.0254e-03 & 2.6150e-02 \\
Q1 & 4.8151 & 196.1414 & 250.0693 & 2.34e-02 & 3.2367e-04 & 3.2367e-04 & 3.2367e-04 & 9.1909e-03 \\
Q2 & 4.6097 & 200.042 & 185.2273 & 7.30e-03 & 2.3726e-04 & 2.3726e-04 & 2.3726e-04 & 1.0677e-02 \\
Q3 & 4.8755 & 191.0563 & 245.5172 & 2.24e-02 & 2.5858e-04 & 2.5858e-04 & 2.5858e-04 & 6.6547e-03 \\
Q4 & 4.8181 & 145.6047 & 193.6868 & 1.62e-02 & 1.7959e-04 & 1.7959e-04 & 1.7959e-04 & 5.0011e-03 \\
Q5 & 4.7342 & 342.2608 & 218.3708 & 5.30e-03 & 1.6729e-04 & 1.6729e-04 & 1.6729e-04 & 6.1321e-03 \\
Q6 & 4.8762 & 396.4902 & 235.0054 & 2.15e-02 & 1.4438e-04 & 1.4438e-04 & 1.4438e-04 & 4.6050e-03 \\
Q7 & 4.9675 & 163.9886 & 164.5995 & 1.80e-02 & 2.1944e-04 & 2.1944e-04 & 2.1944e-04 & 5.8476e-03 \\
Q8 & 4.9024 & 316.3906 & 202.2912 & 9.70e-03 & 1.6401e-04 & 1.6401e-04 & 1.6401e-04 & 2.2242e-02 \\
\end{tabular}
\end{ruledtabular}
\end{table*}

\begin{table*}
\caption{\label{tab:qubit_params_2} Calibration data for the IBM quantum computer {\it ibm\_brisbane} at December 4, 2023, 14:54 PM, Israel time \cite{IBMQuantum}. This calibration matches the time of running and obtaining the results for the feedforward unoptimized circuit plotted at Fig.~\ref{fig_imaginary_time}(b3). We used 11 qubits and showing the calibration for them only.}
\begin{ruledtabular}
\begin{tabular}{ccccccccc}
Qubit name & Frequency [GHz] & T1 [us] & T2 [us] & Readout error & ID error & $\sqrt{X}$ error & Pauli-X error & Next neighbor ECR error\\
Q0 & 4.7219 & 297.1675 & 83.6881 & 2.24e-02 & 1.7105e-04 & 1.7105e-04 & 1.7105e-04 & 7.2451e-03 \\
Q1 & 4.8151 & 314.4306 & 276.3092 & 2.28e-02 & 4.0617e-04 & 4.0617e-04 & 4.0617e-04 & 2.1066e-02 \\
Q2 & 4.6097 & 245.1167 & 233.7216 & 1.18e-02 & 3.2942e-04 & 3.2942e-04 & 3.2942e-04 & 7.8413e-03 \\
Q3 & 4.8756 & 318.1397 & 243.1388 & 2.07e-02 & 1.5340e-04 & 1.5340e-04 & 1.5340e-04 & 3.7432e-03 \\
Q4 & 4.8181 & 226.5137 & 234.2866 & 1.24e-02 & 1.3296e-04 & 1.3296e-04 & 1.3296e-04 & 6.4607e-03 \\
Q5 & 4.7342 & 102.9683 & 165.5786 & 1.65e-02 & 2.6254e-04 & 2.6254e-04 & 2.6254e-04 & 8.2273e-03 \\
Q6 & 4.8762 & 367.6151 & 166.9219 & 9.70e-03 & 1.3747e-04 & 1.3747e-04 & 1.3747e-04 & 4.2064e-03 \\
Q7 & 4.9675 & 215.6241 & 284.9961 & 1.79e-02 & 2.5122e-04 & 2.5122e-04 & 2.5122e-04 & 5.4073e-03 \\
Q8 & 4.9024 & 345.8321 & 171.5487 & 7.20e-03 & 1.5693e-04 & 1.5693e-04 & 1.5693e-04 & 5.7560e-03 \\
Q9 & 4.9872 & 336.6973 & 196.2205 & 6.30e-03 & 1.4140e-04 & 1.4140e-04 & 1.4140e-04 & 9.1016e-03 \\
Q10 & 4.8315 & 168.2043 & 188.116 & 1.67e-02 & 3.2911e-04 & 3.2911e-04 & 3.2911e-04 & 7.6139e-03 \\

\end{tabular}
\end{ruledtabular}
\end{table*}

The calibration data of the IBM quantum computer {\it ibm\_brisbane} are shown in Tables~\ref{tab:qubit_params_1} and \ref{tab:qubit_params_2} for the corresponding optimized and feedforward quantum demonstration plotted in Fig.~\ref{fig_imaginary_time}. The optimized and feedforward quantum demonstrations were done at November 14, 2023, 02:00 AM and December 4, 2023, 14:54 PM Israel time respectively.

\subsubsection{Quantum computer demonstration \& results}

\begin{figure}[t]
\includegraphics[width=\linewidth]{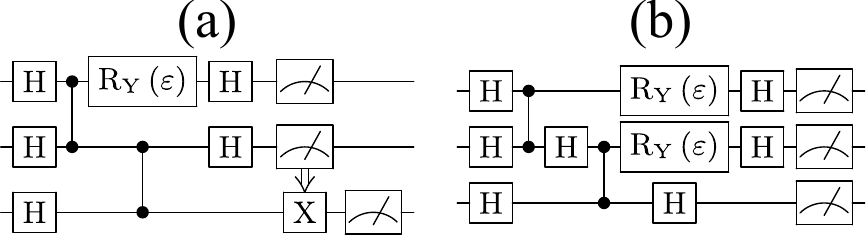}
\caption{\label{fig:circuit_cond_5} (a) Circuit implementation of the gate $\hat{N}_\varepsilon$ of the main text. The $X$ byproduct is canceled by a feed-forward classical conditional gate if the middle qubit is measured $1$ as indicated by the arrow. The input state $\ket{+}$ converges to $\ket{0}$ after applying the same gate multiple times. The output is measured in the $Z$ basis. (b)  Circuit implementation after optimization for NISQ computers, which is equivalent to (a) but uses no feed-forward.}
\end{figure}

To check our protocol robustness to noises and present its operation, we implement and run it on a real quantum computer. We use the Qiskit quantum computation library \cite{Qiskit}. Though the proposed MBQC circuit is dominated by noise, by simple optimizations we overcome the noise and have a striking agreement with the ideal case. We proceed to explain the  circuits and their results.

First, we implement the MBQC protocol as described in the main text. The circuit is shown in Fig.~\ref{fig:circuit_cond_5}(a). This circuit relies only on measurements of the cluster state after incorporating an input state, which is $\ket{\mathrm{in}} = \ket{+}$. This protocol consists of 2 rounds of $\varepsilon=0.25$ non-unitary step operations. Post-selecting on all 4 bulk measurements is one possibility, but we choose to avoid post-selection on the $X$ measurements by feed-forward correction. Therefore, based on the classical result we apply additional $Y$ measurements that cancel the byproduct operators. However, this feed-forward seems to cause severe deterioration of the results which we optimize next.

The optimized circuit is shown in Fig.~\ref{fig:circuit_cond_5}(b). This circuit has each 2-qubit step compacted into 1-qubit by adding an Hadamard gate to perform $H M_{s_1}$ where $s_1$ is the measurement result of the corresponding qubit. Hence, if we apply $H$ on the logical state just after the action of the measurement, we effectively apply $M_{s_1}$, thus we add $H$ between  subsequent CZs. This not only saves 1-qubit per step, but reduces the use of post-selection and feed-forward correction methods, thus, reducing the noise levels dramatically.

We run both circuits on {\it ibm\_ brisbane} and obtain the results shown in Fig.~\ref{fig_imaginary_time}(b3). We choose $\varepsilon = 0.25$ and we vary the number of steps from $n=2$ to $n_{\mathrm max} = 8$, where $n=1$ feedforward circuit is shown in Fig.~\ref{fig:circuit_cond_5}(a) and optimized version for $n=2$ is shown in Fig.~\ref{fig:circuit_cond_5}(b). We plot the probability to measure output of $\ket{0}$, which increases as we increase the number of steps. We plot it for both the ideal case, a noisy simulator provided by Qiskit that mimics the actual noise on the hardware and {\it ibm\_ brisbane} with both optimized and unoptimized circuits. We used 40,000 measurements in all runs except for the real quantum computer where we were limited to 20,000 measurements. Then, we post-selected the $\varepsilon$ angle measurements to $0$. The error bars are calculated from the standard error $\sqrt{\frac{p(1-p)}{n}}$ of a binomial distribution, where $n$ is the number of post-selected measurements and $p=p(0)$. This assumes only statistical noise. Though the ideal and noisy simulator cases show similar results, the unoptimized circuit suffers from too much noise already at $n=3$, showing that the feed-forward protocol is too demanding for real quantum computer and optimization is necessary. A possible explanation for the low fidelity is that conditional gates introduce too much noise. 

As opposed to the feedforward circuit, for the optimized case we reach $n_{\mathrm{max}}=8$ with remarkably high fidelity to the ideal case. We still see some saturation as we increase $n$, but we do not fall to the random case. A possible explanation for the saturation is that the native gates of this quantum computer, which are not $H$ and control-$Z$, require too many gate decompositions. The larger error bars may be explained by the post-selection as its accuracy decreases as $n$ grows. Therefore, our protocol works perfectly as expected with very low amount of noise even on NISQ computers after this simple optimization.

\section{Operator entanglement}

We now detail our measurements of the ``operator entanglement'' of an MBQC action. As discussed in the main text, it is clear that entanglement between the output and the input is necessary for manifesting an MBQC action. The ``operator entanglement'' allows to quantify how ``nonunitary" a gate is.

Since measuring the entanglement entropy $S_{\mathrm{EE}}$ is considered challenging due to its non-polynomial nature, we instead focus on  the 2nd \renyi entanglement $S_2^{\mathrm{op}} = -\log \left( \sum_i \mu_i^4 \right)$, where the $\mu_i$'s are the eigenvalues of $\hat{N}$. 

To measure this quantity on reduced density matrices one usually uses the so-called SWAP test \cite{cincio2018learning,azses2020identification}. As explained in the main text and in Figs.~2(a2,a3), we construct a  Bell state $ \left( \ket{00} + \ket{11} \right)\sqrt{2}$  and apply the nonunitary gate on one of the aubits yielding $\ket{\psi} \propto \sum_i \ket{i}  \otimes \hat{N}\ket{i} = \sum_i \mu_i \ket{i} \otimes \ket{i}$. The overall circuit of one-copy is shown in Figs.~2(a2,a3) of the main text. Therefore, by measuring the \renyi entanglement of this state on either qubit one measures the \renyi operator entanglement.

We present the results for the SWAP protocol for the ideal case in Fig.~\ref{fig_imaginary_time}(a1), notated `SWAP Test'. The data point and the error bars are calculated from the averages and standard deviation of repeating 10 times the execution of the circuit with 20,000 measurements each, which are filtered by post-selection. The \renyi operator entanglement is measured as a function of $\varepsilon$. At the left side, $\varepsilon=0$ and we get a unitary $\hat{N}$ that all its eigenvalues are equal and normalized $\sum_i |\mu_i|^2=1$, thus, $S_2^{\mathrm{op}} = \log(2)$. However, as $\varepsilon$ approaches $\pi/2$ we have more non-unitary projective nature. For $\varepsilon = \pi/2$ w.l.o.g. $\mu_i = 0$ except $\mu_1 = 1$, thus, $S_2^{\mathrm{op}} = 0$. This is shown in the plot as we expect.

A different approach for measuring the \renyi entanglement which saves quantum resources is the randomized measurements approach \cite{elben2019statistical,elben2022randomized,zhou2017operator}. In this approach, one first measures in many random bases and only then calculates the desired quantity by classical post-processing \cite{huang2020predicting}. We use here Haar random unitaries as provided by Qiskit on the first qubit, which is the MBQC output, in Fig.~\ref{fig_imaginary_time}(a2). We run the protocol 10 times with $N=40$ random unitaries and $K=500$ measurements each on an ideal simulator. Here we use the classical post-processing protocol that employs the identity \cite{elben2018renyi,vermersch2018unitary,elben2019statistical}
\be
S_2^{\mathrm{op}} = -\log \left[ 2 \sum_{s,s'} (-2)^{-D(s,s')} P(s) P(s') \right],
\ee
where $s$ and $s'$ are the MBQC output qubit measurement result, $P(s)$ and $P(s')$ are averaged over all the random unitaries, and $D(s,s')=1$ if $s\neq s'$ and $0$ otherwise. The data point and its error bars are derived from the average and standard deviation of the 10 runs. We notate this approach as `Hamming Formula' in the main text Fig.~\ref{fig_imaginary_time}(a1). 

The advantage of the random measurements scheme is that it allows the efficient extraction of more than one quantity by employing random tomography and classical shadow copies techniques \cite{huang2020predicting}. Mainly, after the measurement of the random unitary, one inverses the quantum channel using \cite{elben2022randomized}
\be
\bar{\rho}_m = 3 U_m^\dagger \left[\frac{1}{K} \sum_{i=1}^K\ket{s_i}\bra{s_i} \right] U_m - I,
\ee
where $U_m$ is the random unitary, $K=500$ is the number of measurements for each random unitary and $\ket{s_i}$ is the measurement result of the output qubit. We obtain $M=40$ such reduced density matrices.
Then, we calculate $S_2^{\mathrm{op}}$ using the equation
\be
S_2^{\mathrm{op}} = -\log \left[ \frac{1}{M(M-1)} \sum_{m\neq m'}\Tr( \bar{\rho}_m \bar{\rho}_m')\right].
\ee
We repeat this for 10 times on an ideal simulator.
Moreover, one may not need to choose random Haar unitaries as the random Clifford group is 3-design \cite{webb2016clifford} and has efficient sampling method \cite{bravyi2021hadamard}, thus, it allows the perfect estimation of even the 3rd \renyi operator entanglement \cite{vermersch2018unitary}. Large error bars arise in our estimations for both random measurements methods may be due to the highly entangled to lower entangled transition, as the two regimes require different ratio of random unitaries to the number of random measurements \cite{elben2018renyi, brydges2019probing}. The results are shown in Fig.~\ref{fig_imaginary_time}(a1). We notate this approach as `Random Tomography' in the main text Fig.~\ref{fig_imaginary_time}(a1). The data point and its error bars are derived from the average and standard deviation of the 10 runs. As we can see, both randomized measurement results match the previous more quantum resource intensive SWAP test method, although with larger error bars.

\end{appendix}

\end{document}